\documentclass[%
 reprint,
 amsmath,amssymb,
 aps,
prd,
]{revtex4-1}

\usepackage{graphicx}
\usepackage{dcolumn}
\usepackage{bm}

\begin{document}

\preprint{APS/123-QED}

\title{Joule-Thomson Expansion of Charged AdS Black Holes}
\author{\"{O}zg\"{u}r \"{O}kc\"{u}}
\email{ozgur.okcu@ogr.iu.edu.tr}
\author{Ekrem Ayd{\i}ner}

 \email{ekrem.aydiner@istanbul.edu.tr}
\affiliation{Department of
Physics, Faculty of Science, \.{I}stanbul University,
\.{I}stanbul, 34134, Turkey}%

\date{\today}

\begin{abstract}
 In this paper, we study Joule-Thomson effects for charged AdS black holes. We obtain inversion temperatures and curves. We investigate  similarities and differences between van der Waals fluids and charged AdS black holes for the expansion. We obtain isenthalpic curves for both systems in $T-P$ plane and determine the cooling-heating regions.
\end{abstract}

\pacs{04.70.-s, 05.70.Ce, 04.60.-m}
\maketitle


\section{Introduction}
It is well known that black holes as thermodynamic systems have many interesting consequences. It sets deep and fundamental connections between the laws of classical general relativity, thermodynamics, and quantum mechanics. Since it has a key feature to understand quantum gravity, much attention has been paid to the topic. The properties of black hole thermodynamics have been investigated since the first studies of Bekenstein and Hawking \cite{Bek1972,Bek1973,Bar1973,Bek1974,Haw1974,Haw1975}. When Hawking discovered that black holes radiate, black holes are considered as thermodynamic systems.

Black hole thermodynamics shares similarities with general thermodynamics systems. Specifically, black holes in AdS space have common properties with general systems. The study of AdS black hole thermodynamics began with pioneering paper of Hawking and Page \cite{Haw1983}. They found a phase transition between Schwarzschild AdS black hole and thermal AdS space. Up to now, thermodynamic properties of AdS black holes have been widely studied in the literature \cite{Cham1999a,Cham1999b,Niu2012,Tsai2012,Bane2011,Bane2012,Cald2000,Kas2009,Kub2012,Dolan2011,Beh2013,Cai2013,Belhaj2015,Guna2012,Dutta2013,Li2014,Alta2014,Liang2016,Hendi2013,Hendi2016,Spa2013,Moa2013,Alta2014b,Sadeghi2014,Zhao2013,Dolan2011b,John2014,John2015a,John2015b,John2016,Bel2015b,Caceres2015,Sadeghi2015,Setare2015}. In \cite{Cham1999a,Cham1999b}, authors studied the thermodynamics of charged AdS black holes and they found analogy between phase diagrams of black hole and van der Waals fluids. When cosmological constant and its conjugate quantity are, respectively, considered as a thermodynamic pressure
\begin{equation}
\label{press.}
P=-\frac{\Lambda}{8 \pi},
\end{equation}
and thermodynamic volume $V=(\frac{\partial M}{\partial P})_{S,Q,J}$, this analogy gains more physical meaning. Particularly, in the extended phase space (including P and V terms in the first law of black hole thermodynamics), charged AdS black holes phase transition is remarkable coincidence with van der Waals liquid-gas phase transition \cite{Kub2012}. This type of transition is not limited with charged AdS black holes, various kind of black holes in AdS space show the same phase transitions \cite{Dolan2011,Beh2013,Cai2013,Belhaj2015,Guna2012,Dutta2013,Li2014,Alta2014,Liang2016,Hendi2013,Hendi2016}.

It is also possible to consider heat cycle for AdS black holes \cite{John2014,John2015a,John2015b,John2016,Bel2015b,Caceres2015,Sadeghi2015,Setare2015}. In \cite{John2014,John2016}, author suggested two kind of heat cycles and obtained exact efficiency formula for black holes.

Variable cosmological constant notion has some nice features such as phase transition, heat cycles and compressibility of black holes \cite{Dolan2011b}. Applicabilities of these thermodynamic phenomena to black holes encourage us to consider Joule-Thomson expansion of charged AdS black holes. In this letter, we study the Joule-Thompson expansion for chraged AdS black holes. We find similarities and differences with van der Waals fluids. In  Joule-Thomson expansion, gas at a high pressure passes through a porous plug to a section with a low pressure and during the expansion enthalpy is constant. With the Joule-Thomson expansion, one can consider heating-cooling effect and inversion temperatures.

The paper arranged as follows. In Section \ref{CABH}, we briefly review the charged AdS black hole. In Section \ref{JTE}, we firstly review Joule-Thompson expansion for van der Waals gases and then we investigate Joule Thomson expansion  for charged AdS black holes. Finally, we discuss our result in Section \ref{Res}. (Here we use the units $G_{N}=\hbar=k_{B}=c=1.$)

\section{Charged AdS Black Holes}
\label{CABH}
In this section, we briefly review charged AdS black hole and we present its thermodynamic properties. 
Charged black hole in four dimensional space is defined with the metric
\begin{equation}\label{met}
ds^{2}
=-f(r)dt^{2}
+f^{-1}(r)
dr^{2}
+r^{2}
d\Omega^{2},
\end{equation}
where $d\Omega^{2}=d\theta^{2}+\sin^{2}(\theta)d\phi^{2}$
and $f(r)$ is given by
\begin{equation}\label{metf}
f(r)=1-\frac{2M}{r}
+\frac{Q^{2}}{r^{2}}
+\frac{r^{2}}{l^{2}} \ .
\end{equation}
In these equations, $l$, $M$ and $Q$ are the AdS radius, mass, and charge of the black hole, respectively. One can obtain black hole event horizon as largest root of $f(r_{+})=0$. The mass of black hole in Eq.\,(\ref{metf}) is given by
\begin{equation}\label{mass}
M=\frac{r_{+}}{2}
\left(1+\frac{Q^{2}}{r_{+}^{2}}
+\frac{r_{+}^{2}}{l^{2}} \right)
\end{equation}
which satisfies the first law of black hole thermodynamics
\begin{equation}\label{flbt}
dM=T dS+\Phi dQ+V dP \
\end{equation}
and corresponding Smarr relation is given by
\begin{equation}\label{smarr}
M=2(TS-PV)+\Phi Q \ .
\end{equation}
One can derive Smarr relation by scaling argument \cite{Kas2009}. The first law of black hole thermodynamic includes $P$ and $V$, when the cosmological constant is considered as a thermodynamic variable. The cosmological constant corresponds to the pressure,
\begin{equation}\label{press2}
P=-\frac{1}{8\pi}
\Lambda
=\frac{3}{8\pi}
\frac{1}{l^{2}}
\end{equation}
and cosmological constant's conjugate quantity corresponds to thermodynamic volume. The expression for entropy is given by
\begin{equation}\label{ent}
S=\frac{A}{4}=\pi r_{+}^{2},\quad A=4\pi r^{2}_{+}
\end{equation}
and the corresponding Hawking temperature
\begin{equation}\label{temp}
T=\left(\frac{\partial M}{\partial S}
\right)_{P,Q}
=\frac{l^{2}(r_{+}^{2}-Q^{2})+3r_{+}^{4}}{4\pi l^{2} r_{+}^{3}} \ .
\end{equation}
On the other hand, the electric potential is given by
$\Phi=\frac{Q}{r_{+}}$ 
and equation of state $P=P(V,T)$ for charged AdS black hole is obtained from Eqs.\,(\ref{press2}) and (\ref{temp}) as
\begin{equation}\label{eos}
P=\frac{T}{2r_{+}}
-\frac{1}{8\pi r_{+}^{2}}
+\frac{Q^{2}}{8\pi r_{+}^{4}}, \quad r_{+}
=(\frac{3V}{4\pi}
)^{\frac{1}{3}}\, .
\end{equation}
The critical points \cite{Kub2012} obtained from
\begin{equation}\label{crp}
\frac{\partial P}{\partial r_{+}}=0,\quad
\frac{\partial^{2}P}{\partial r_{+}^{2}}=0, 
\end{equation}
which leads to
\begin{equation}\label{crp1}
T_{c}
=\frac{\sqrt{6}}{18\pi Q}, \quad
r_{c}
=\sqrt{6} Q, \quad
P_{c}
=\frac{1}{96\pi Q^{2}} \ .
\end{equation}
Other thermodynamic properties can be obtained by using above relations. For example, heat capacities at constant pressure and constant volume are, respectively, given by
\begin{equation}\label{Cp}
C_{P}
=T\left(\frac{\partial S}{\partial T}\right)_{P,Q}=2\pi
r^{2}
\frac{3r_{+}^{4}-l^{2}Q^{2}+l^{2}r_{+}^{2}}{3r_{+}^{4}+3l^{2}Q^{2}-l^{2}r_{+}^{2}} \ ,
\end{equation}
and
\begin{equation}\label{Cv}
C_{V}=T\left(\frac{\partial S}{\partial T}\right)_{V,Q}=0 \ .
\end{equation}

In this section, we give thermodynamic definitions for charged AdS black hole. In the next section, we will review  Joule-Thomson expansion for van der Waals fluids and investigate Joule-Thomson expansion for charged AdS black holes.

\section{Joule-Thomson Expansion}
\label{JTE}
In this section, we review the well-known Joule-Thomson expansion \cite{Wint,Johns2014}. In Joule-Thomson expansion, gas at a high pressure passes through a porous plug or  small valve to a section with a low pressure in a thermally insulated tube and enthalpy remains constant during the expansion process. One can describe temperature change with respect to pressure and this change is given by

\begin{equation}\label{JT1}
\mu=\left(\frac{\partial T}{\partial P}\right)_{H} \ .
\end{equation}
Here $\mu$ is called the Joule-Thomson coefficient. It is possible to determine whether cooling or heating will occur by evaluating the sign of Eq.\,(\ref{JT1}). In Joule-Thomson expansion, pressure decreases so change of pressure is negative but change of temperature may be positive or negative. If the change of temperature is positive (negative) $\mu$ is negative (positive) and so gas warms (cools).

It is also possible to express Eq.\,(\ref{JT1}) in terms of volume and heat capacity at constant pressure. From the first law of thermodynamics, one can write the fundamental relation for constant particle number $N$
\begin{equation}\label{firstLawGeneral}
dU=TdS-PdV \ .
\end{equation}
Using the relation $H=U+PV$, Eq.\,(\ref{firstLawGeneral}) is given by
\begin{equation}\label{firstLawGeneralH}
dH=TdS+VdP \ .
\end{equation}
Since $dH=0$, Eq.\,(\ref{firstLawGeneralH}) is given by
\begin{equation}\label{enthalph2}
0=T\left(\frac{\partial S}{\partial P}\right)_{H}+V \ .
\end{equation}
Since entropy is a state function, the differential $dS$ is given by
\begin{equation}\label{entrop}
dS=\left(\frac{\partial S}{\partial P}\right)_{T}dP+\left(\frac{\partial S}{\partial T}\right)_{P}dT \ 
\end{equation}
which can be rearranged to give
\begin{equation}\label{entrop2}
\left(\frac{\partial S}{\partial P}\right)_{H}=\left(\frac{\partial S}{\partial P}\right)_{T}+\left(\frac{\partial S}{\partial T}\right)_{P}\left(\frac{\partial T}{\partial P}\right)_{H} \ .
\end{equation}
If one can substitute this expression into Eq.\,(\ref{enthalph2}), one can obtain the following expression:
\begin{equation}\label{entrop3}
0=T\left[\left(\frac{\partial S}{\partial P}\right)_{T}+\left(\frac{\partial S}{\partial T}\right)_{P}\left(\frac{\partial T}{\partial P}\right)_{H}\right]+V \ .
\end{equation}
Substituting Maxwell relation $\left(\frac{\partial S}{\partial P}\right)_{T}=-\left(\frac{\partial V}{\partial T}\right)_{P}$ and $C_{P}=T\left (\frac{\partial S}{\partial T}\right)_{P}$ into Eq.\,(\ref{entrop3}) gives
\begin{equation}\label{entrop4}
0=-T\left(\frac{\partial V}{\partial T}\right)_{P}+C_{P}\left(\frac{\partial T}{\partial P}\right)_{H}+V \ 
\end{equation}
and it can be rearranged to give the Joule-Thomson coefficient \cite{Wint} as follows:
\begin{equation}\label{JT2}
\mu=\left(\frac{\partial T}{\partial P}\right)_{H}=\frac{1}{C_{P}}\left[T\left(\frac{\partial V}{\partial T}\right)_{P}-V\right] \ .
\end{equation}
At the inversion temperature, $\mu$ equals zero and inversion temperature is given by
\begin{equation}\label{iT}
T_{i}=V\left(\frac{\partial T}{\partial V}\right)_{P} \ 
\end{equation}
which is useful to determine the heating and cooling regions in the $T-P$ plane.

\subsection{van der Waals Fluids}
The van der Waals equation is a generalized form of ideal gas equation, which usually describes the liquid-gas phase transition behaviours for real fluids \cite{Johns2014,Vent2001}. It takes into account the size of molecules and attraction between them. 
It is given by 
\begin{equation}\label{vdW}
P=\frac{k_{B}T}{v-b}-\frac{a}{v^{2}} \ .
\end{equation}
Here $v=\frac{V}{N}$, $P$, $T$ and $k_{B}$ denote the specific volume, pressure, temperature, and Boltzmann constant. $a>0$ constant is a measure of attraction between particles and $b>0$ is a measure of molecule volume.  $a$ and $b$ constants are determined from experimental data.

Before more proceeding to the Joule-Thomson expansion, it is useful to give some thermodynamic properties of van der Waals equation. Following \cite{Kub2012,Golden}, free energy is given by
\begin{equation}\label{fe}
F(T,v)=-k_{B}T\left(1+ln\left[\frac{(v-b)T^{\frac{3}{2}}}{\Phi}\right]\right)-\frac{a}{v} \ .
\end{equation}
Here $\phi$ is a constant characterizing the gas. Now, entropy can be obtained from Eq.\,(\ref{fe})
\begin{equation}\label{entvdW}
S(T,v)=-\left(\frac{\partial F}{\partial T}\right)_{v}=k_{B}\left(\frac{5}{2}+ln\left[\frac{(v-b)T^{\frac{3}{2}}}{\Phi}\right]\right) \ .
\end{equation}
Using Eqs.\, (\ref{fe}) and (\ref{entvdW}), we can calculate the internal energy
\begin{equation}\label{ie}
U(T,v)=F+TS=\frac{3k_{B}T}{2}-\frac{a}{v} \ 
\end{equation}
and from Eqs.\,(\ref{vdW}) and (\ref{ie}), enthalpy is
\begin{equation}\label{enthalphy}
H(T,v)=U+PV=\frac{3k_{B}T}{2}+\frac{k_{B}Tv}{v-b}-\frac{2a}{v} \ .
\end{equation}

\begin{figure}[h!]
		\centering
		\includegraphics{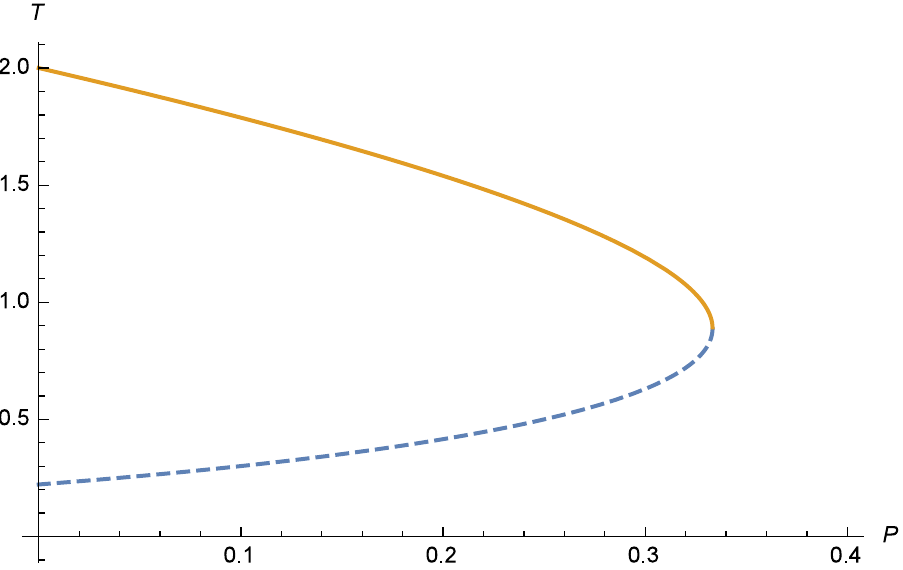}
	\caption{Lower (dashed blue line) and upper (solid orange line) inversion curves. We fix $a=b=k_{B}=1$.}
	\label{iC}
\end{figure}

Now, let us calculate the inversion temperature for van der Waals equation. Using Eq.\,(\ref{iT}), inversion temperature is given by
\begin{equation}\label{ie2}
T_{i}=\frac{1}{k_{B}}\left(P_{i}v-\frac{a}{v^{2}}(v-2b)\right) \ 
\end{equation} 
where $P_{i}$ denotes the inversion pressure. From Eq.\,(\ref{vdW}), one can get
\begin{equation}\label{ie3}
T_{i}=\frac{1}{k_{B}}\left(P_{i}+\frac{a}{v^{2}}\right)(v-b) \ . 
\end{equation}
Subtracting Eq.\,(\ref{ie2}) from Eq.\,(\ref{ie3}) yields
\begin{equation}\label{equa}
bP_{i}v^{2}-2av+3ab=0 \ 
\end{equation}
and, solving this equation for $v(P_{i})$, one can obtain two roots
\begin{equation}\label{roots}
v=\frac{a\pm\sqrt{a^{2}-3ab^{2}P_{i}}}{bP_{i}} \ .
\end{equation}
Substituting these roots into Eq.\,(\ref{ie3}), one can obtain
\begin{equation}\label{lower}
T^{lower}_{i}=\frac{2\left(5a-3b^{2}P_{i}-4\sqrt{a^{2}-3ab^{2}P_{i}}\right)}{9bk} \ ,
\end{equation}
\begin{equation}\label{upper}
T^{upper}_{i}=\frac{2\left(5a-3b^{2}P_{i}+4\sqrt{a^{2}-3ab^{2}P_{i}}\right)}{9bk} \ 
\end{equation}
which give lower and upper inversion curves, respectively. In Fig.\,(\ref{iC}), lower and upper inversion curves are presented. At the point $P_{i}=0$, we can obtain the minimum and maximum inversion temperatures
\begin{equation}\label{T0}
T_{i}^{min}=\frac{2a}{9bk}, \qquad T_{i}^{max}=\frac{2a}{bk} \ .
\end{equation}
The critical temperature for van der Waals fluids is given by $T_{c}=\frac{8a}{27bk}$ and hence
\begin{equation}\label{TRatios}
\frac{T_{i}^{min}}{T_{c}}=\frac{3}{4}, \qquad \frac{T_{i}^{max}}{T_{c}}=\frac{27}{4} \ .
\end{equation}

\begin{figure}
		\centering
		\includegraphics{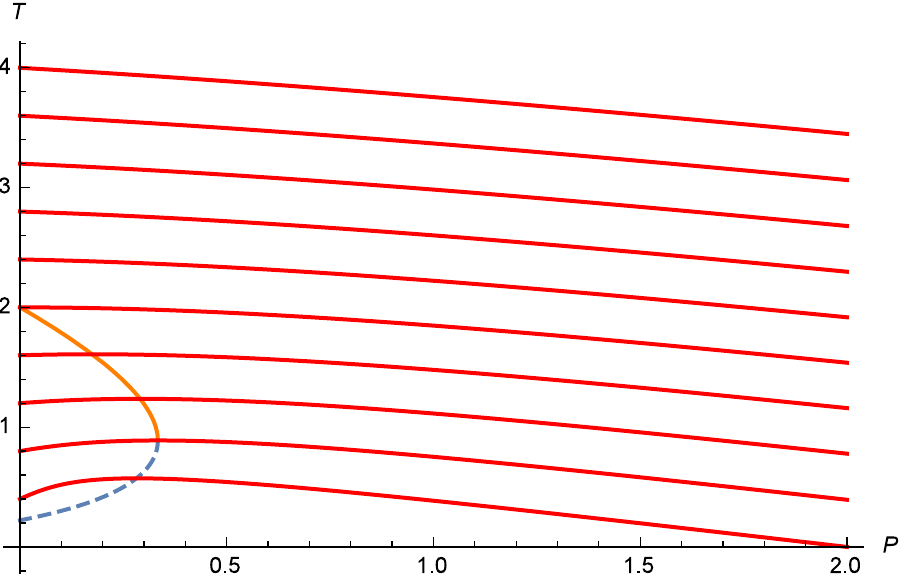}
	\caption{Dashed blue line and orange line are inversion curves. Red lines are isenthalpic curves. The enthalpies of isenthalpic curves increase from bottom to top and  correspond to $H=1,2,3,4,5,6,7,8,9,10$. We fix $a=b=k_{B}=1$.}
	\label{iE}
\end{figure}

Using Eqs.\,(\ref{vdW}) and (\ref{enthalphy}), we can obtain the isenthalpic curves in $T-P$ plane. In Fig.\,(\ref{iE}), isenthalpic and inversion curves are presented. When the isenthalpic curves cross inversion curves, their slopes change sign. Isenthalpic curves have positive slopes inside the inversion curves, otherwise their slopes are negative. As a result the Joule-Thomson coefficient is positive inside the inversion curves and cooling occurs inside this region.

\subsection{Charged AdS Black Holes}
\begin{figure}
		\centering
		\includegraphics{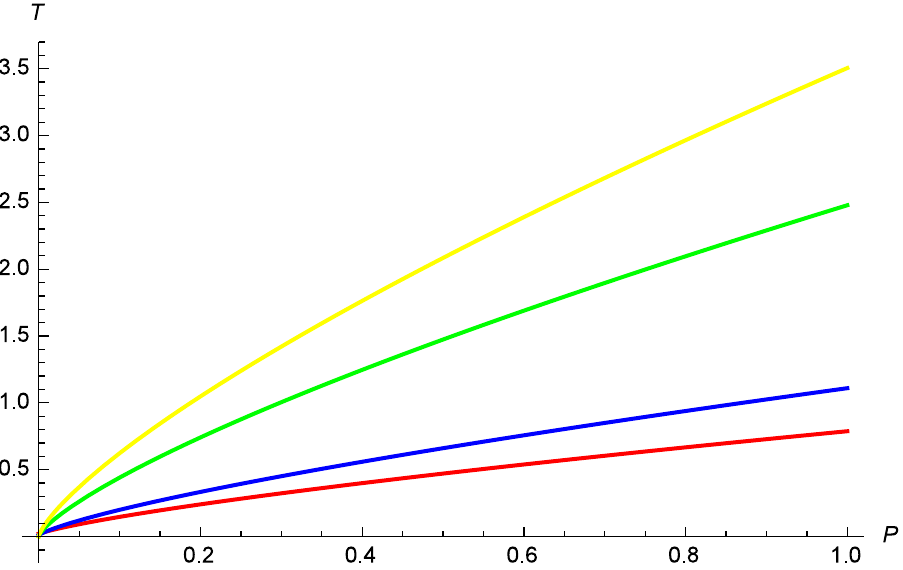}
	\caption{Inversion curves for charged AdS black hole.From bottom to top, the curves correspond to $Q=1,2,10,20$.}
	\label{iC2}
\end{figure}

In this section, we will consider Joule-Thomson expansion for charged AdS black holes. In \cite{Kas2009}, authors suggested that black hole mass is considered as the enthalphy in AdS space. It means that our isenthalpic curves are actually constant mass curves in AdS space. We can consider black hole mass not to change during the Joule-Thomson expansion. For a fixed charge, similar steps in the previous section can be used to obtain the Joule-Thomson coefficient. Hence

\begin{equation}\label{JTRNAdS}
\mu=\left(\frac{\partial T}{\partial P}\right)_{M}=\frac{1}{C_{P}}\left[T\left(\frac{\partial V}{\partial T}\right)_{P}-V\right] \ .
\end{equation}
The charged AdS black hole equation of state can be given in terms of thermodynamic volume,
\begin{equation}\label{eosRnAdS}
T=\frac{1}{3}\left(\frac{3}{4\pi}\right)^{\frac{2}{3}}V^{\frac{1}{3}}\left[8\pi\left(\frac{3}{4\pi}\right)^{\frac{2}{3}}P+\frac{1}{V^{\frac{2}{3}}}-\left(\frac{4\pi}{3}\right)^{\frac{2}{3}}\frac{Q^{2}}{V^{\frac{4}{3}}}\right]
\ 
\end{equation}
and evaluating this in the right hand side of Eq.\,(\ref{JTRNAdS}), the inversion temperature is given by
\begin{eqnarray}\label{iTRnAdS}
&T_{i}=\frac{1}{3}\left(\frac{6}{\pi}\right)^{\frac{1}{3}}V^{\frac{1}{3}}\left[\left(\frac{\pi}{6}\right)^{\frac{1}{3}}\frac{Q^{2}}{V^{\frac{4}{3}}}-\left(\frac{6}{\pi}\right)^{\frac{1}{3}}\frac{1}{12V^{\frac{2}{3}}}+P_{i}\right]\nonumber\\ &=\frac{Q^{2}}{4\pi r_{+}^{3}}-\frac{1}{12\pi r_{+}}+\frac{2P_{i}r_{+}}{3}
\ .
\end{eqnarray}
From Eq.\,(\ref{eosRnAdS}), one can get
\begin{eqnarray}\label{iTRnAdS2}
&T_{i}=\frac{1}{3}\left(\frac{3}{4\pi}\right)^{\frac{2}{3}}V^{\frac{1}{3}}\left[8\pi\left(\frac{3}{4\pi}\right)^{\frac{2}{3}}P_{i}+\frac{1}{V^{\frac{2}{3}}}-\left(\frac{4\pi}{3}\right)^{\frac{2}{3}}\frac{Q^{2}}{V^{\frac{4}{3}}}\right]\nonumber\\ &=-\frac{Q^{2}}{4\pi r_{+}^{3}}+\frac{1}{4\pi r_{+}}+2P_{i}r_{+}
\ . 
\end{eqnarray}
Subtracting Eq.\,(\ref{iTRnAdS}) form Eq.\,(\ref{iTRnAdS2}) we can obtain

\begin{equation}
8\pi P_{i}r_{+}^{4}+2r_{+}^{2}-3Q^{2}=0
\end{equation}
and solving this equation for $r_{+}$ gives us four roots but only one root is physically meaningful, other roots are complex or negative. A positive and real root is
\begin{equation}\label{root}
r_{+}=\frac{1}{2\sqrt{2}}\sqrt{\frac{\sqrt{1+24P_{i}\pi Q^{2}}}{P_{i}\pi}-\frac{1}{P_{i}\pi}} \ .
\end{equation}
If we substitute this root into Eq.\,(\ref{iTRnAdS2}), the inversion temperature is given by
\begin{equation}\label{iTRnAdS3}
T_{i}=\frac{\sqrt{P_{i}}}{\sqrt{2\pi}}\frac{\left(1+16\pi P_{i}Q^{2}-\sqrt{1+24\pi P_{i}Q^{2}}\right)}{\left(-1+\sqrt{1+24\pi P_{i}Q^{2}}\right)^{\frac{3}{2}}} \ .
\end{equation}
When $P_{i}$ is zero, we have $T^{min}_{i}$
\begin{equation}\label{iTmin}
T^{min}_{i}=\frac{1}{6 \sqrt{6}\pi Q} \ 
\end{equation}
and ratio between minimum inversion and critical temperatures is 
\begin{equation}\label{TRatios2}
\frac{T_{i}^{min}}{T_{c}}=\frac{1}{2} \ .
\end{equation}
In Fig\,(\ref{iC2}), inversion curves are presented for various values of charge $Q$. There is only a lower inversion curve. In contrast to van der Waals fluids, the expression inside the square root in Eq.\,(\ref{iTRnAdS3}) is always positive, so this curve does not terminate any point. 

\begin{figure*}
	\includegraphics{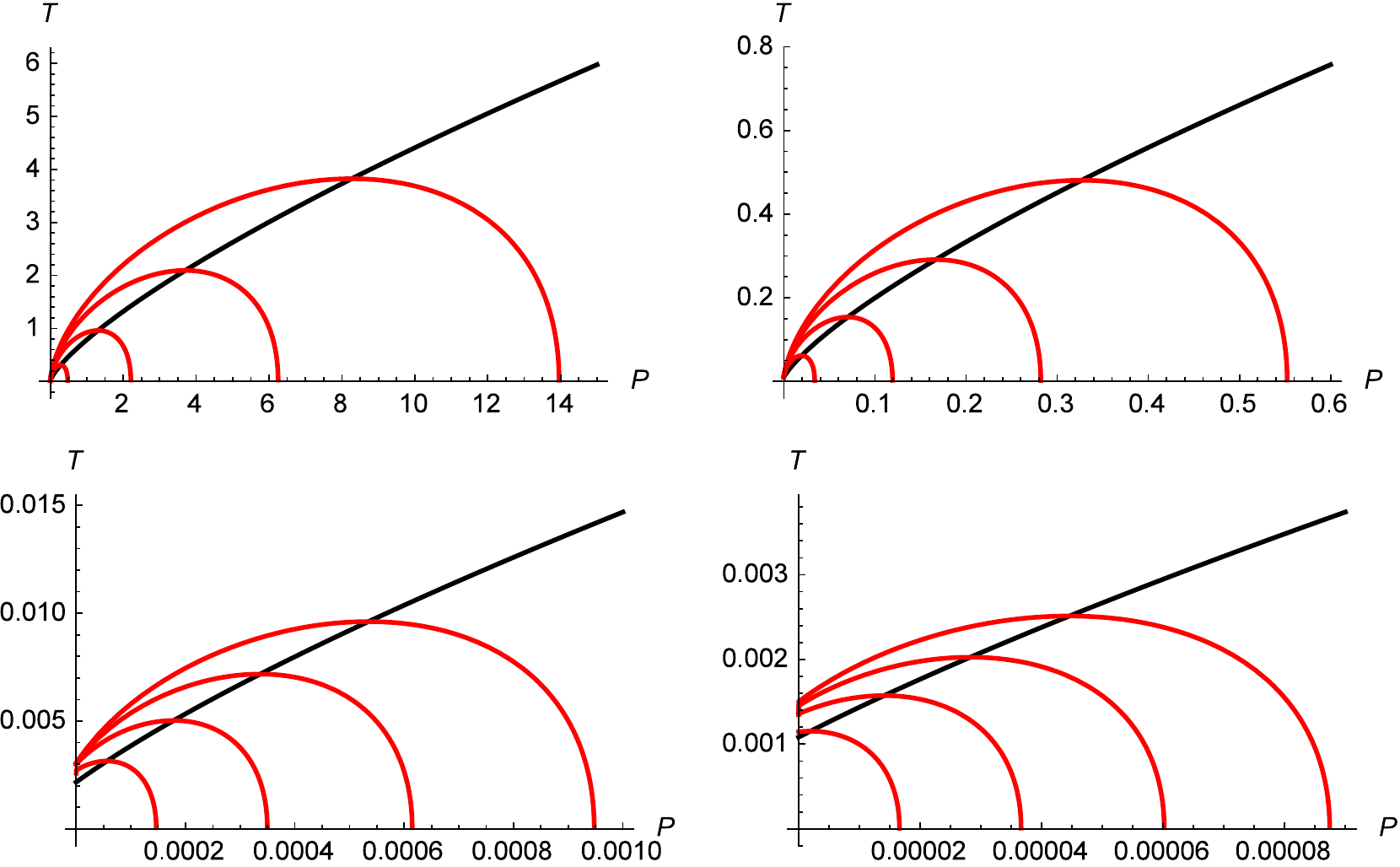}
	\caption{Inversion and isenthalpic curves for charged AdS black hole. From bottom to top, the isenthalpic curves correspond to increasing values of $M$. Red and black lines are isenthalpic and inversion curves, respectively. (a) $Q=1$ and $M=1.5,2,2.5,3$ (b) $Q=2$ and $M=2.5,3,3.5,4$ (c) $Q=10$ and $M=10.5,11,11.5,12$ (d) $Q=20$ and $M=20.5,21,21.5,22$}
	\label{RNiM}
\end{figure*}
\begin{figure*}
	\includegraphics[width=17cm]{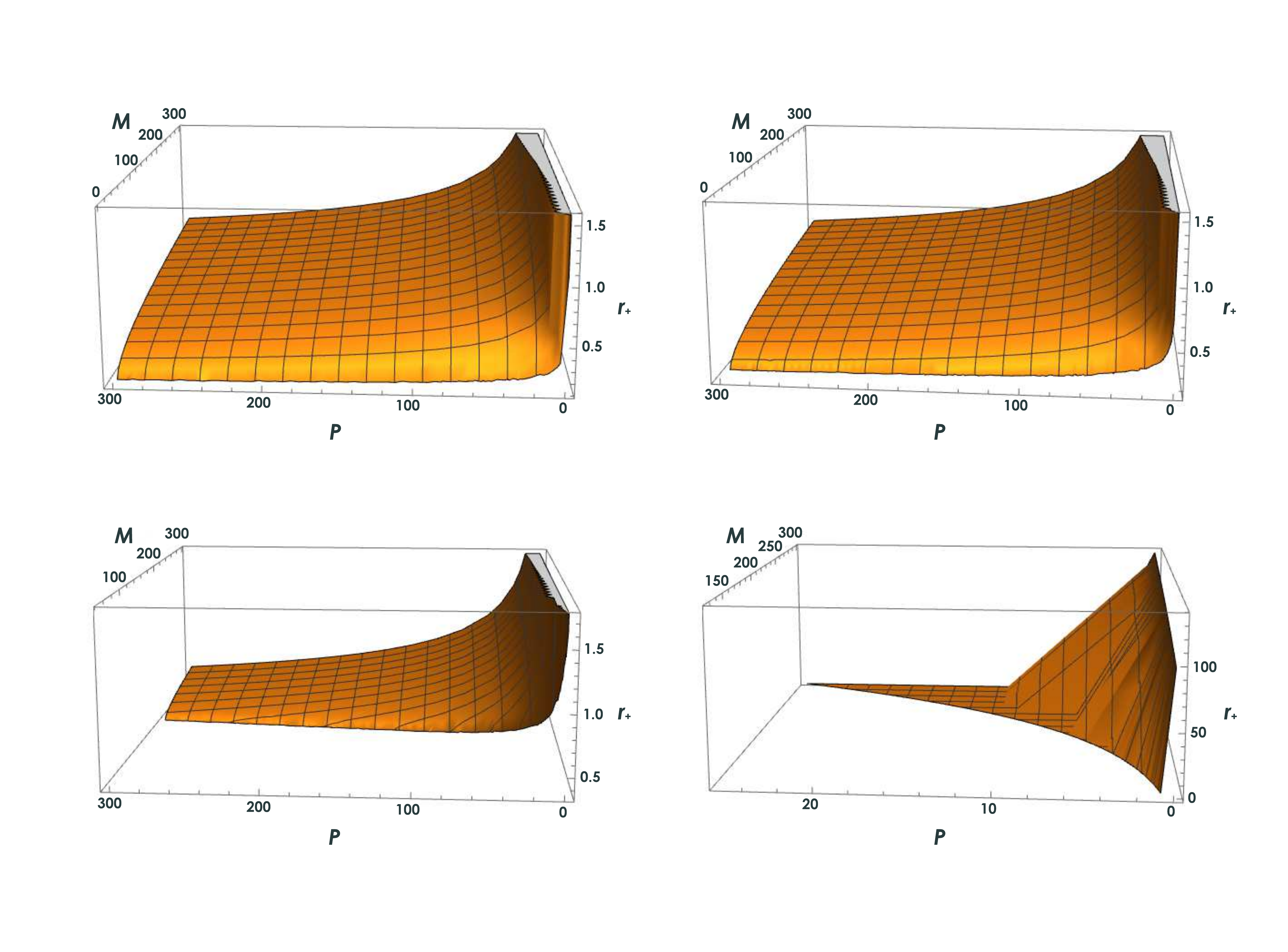}
	\caption{Event horizon of charged AdS black hole. (a) $Q=1$ (b) $Q=2$ (c) $Q=10$ (d) $Q=20$ }
	\label{nakeSing}
\end{figure*}

Now, we can plot isenthalpic, i.e. constant mass, curves in $T-P$ plane. From Eq.\,(\ref{mass}), one can obtain event horizon and substituting event horizon into Eq.\,(\ref{eos}) gives isenthalpic curves in $T-P$ plane. In Fig.\,(\ref{RNiM}), inversion curves and isenthalpic curves are presented. Isenthalpic curves have positive slope above the inversion curves so cooling occurs above the inversion curves. The sign of slope changes under the inversion curves and heating occurs in this region. It is also interesting to talk about naked singularities for charged AdS black holes. In Fig.\,(\ref{nakeSing}), we plot event horizon versus mass and pressure. We introduce four graphics, which correspond to $Q=1, 2, 10, 20.$ The regions can be  seen that denote the naked singularities in Fig.\,(\ref{nakeSing}). One cannot consider Joule-Thomson expansion due to the lack of event horizon for naked singularity. For example, we cannot define event horizon for $Q=20$ and $M\leq20$. For these values, event horizon is imaginary and it corresponds to naked singularity so isenthalpic curves in $T-P$ plane are imaginary. 

\section{Conclusion}
\label{Res}
In this paper, we studied the well known Joule-Thomson expansion for charged AdS black hole. The black hole mass in AdS space is identified with enthalpy due to variable cosmological constant notion, so one can consider that mass does not change during the expansion. First, we reviewed Joule-Thomson expansion for van der Waals fluids and then we investigated Joule-Thomson expansion for charged AdS black holes. We only found one inversion curve that corresponds to the lower curve. It means that black holes always cool above the inversion curve  during the Joule-Thomson expansion.  Cooling and heating regions were shown for various values of charge $Q$ and mass $M$. We also denoted naked singularity which is not sensible for Joule-Thomson expansion due to the lack of event horizon.

Both systems are not well behaved for low temperatures. Unfortunately, isenthalpic curves have positive slopes under the lower inversion curves for both systems. It is also known that van der Waals equation does not too well agree with experiments. Thus Joule-Thomson expansion have been investigated for various equations of state. In charged AdS case, it needs further investigation.
\

\

\begin{acknowledgements}
	We thank Can Onur Keser with improving the figures in this work. This work was sported by Scientific Research Projects Coordination Unit of Istanbul University. Project number is FYL-2016-20615.
\end{acknowledgements}


\clearpage
\begin{references}
\bibitem{Bek1972}
J. D. Bekenstein, Lett. Nuovo Cimento \textbf{4}, 737 (1972)

\bibitem{Bek1973}
J. D. Bekenstein, Phys. Rev. D \textbf{7}, 97 (2333)

\bibitem{Bar1973}
J. M. Bardeen, B. Carter, S. W. Hawking, Commun. Math. Phys. \textbf{31}, 161 (1973)

\bibitem{Bek1974}
J. D. Bekenstein, Phys. Rev. D \textbf{9}, 3292 (1974)

\bibitem{Haw1974}
S. W. Hawking, Nature \textbf{248}, 30 (1974)

\bibitem{Haw1975}
S. W. Hawking, Commun. Math. Phys. \textbf{43}, 199 (1975)

\bibitem{Haw1983}
S. W. Hawking, D. N. Page, Commun. Math. Phys. \textbf{87}, 577 (1983)

\bibitem{Cham1999a}
A. Chamblin, R. Emparan, C. V. Johnson, R. C. Myers, Phys. Rev. D \textbf{60}, 064018 (1999)

\bibitem{Cham1999b}
A. Chamblin, R. Emparan, C. V. Johnson, R.C. Myers, Phys. Rev. D \textbf{60}, 104026 (1999)

\bibitem{Niu2012}
C. Niu, Y. Tian, X. N. Wu, Phys. Rev. D \textbf{85}, 024017 (2012)

\bibitem{Tsai2012}
Y. D. Tsai, X. N. Wu, Y. Yang, Phys. Rev. D \textbf{85}, 044005 (2012)

\bibitem{Bane2011}
R. Banerjee, S. K. Modak, Phys. Rev. D \textbf{84}, 064024 (2011)

\bibitem{Bane2012}
R. Banerjee, S. K. Modak, D. Roychowdhury, J. High Energy Phys. \textbf{10}, 125 (2012)

\bibitem{Cald2000}
M. M. Caldarelli, G. Cognola, D. Klemm, Class. Quant. Grav. \textbf{17}, 399 (2000)

\bibitem{Kas2009}
D. Kastor, S. Ray, J. Traschen, Class. Quant. Grav. \textbf{26}, 195011 (2009)

\bibitem{Kub2012}
D. Kubiznak, R. B. Mann, J. High Energy Phys. \textbf{07}, 033 (2012)

\bibitem{Dolan2011}
B. P. Dolan, Class. Quant. Grav. \textbf{28}, 235017 (2011)

\bibitem{Beh2013}
A. Belhaj, M. Chabab, H. E. Moumni, L. Medari, M. B. Sedra, Chin. Phys. Lett. \textbf{30}, 090402 (2013)

\bibitem{Cai2013}
R. G. Cai, L. M. Cao, L. Li, R. Q. Yang, J. High Energy Phys. \textbf{9}, 005 (2013)

\bibitem{Belhaj2015}
A. Belhaj, M. Chabab, H. E. Moumni, K. Masmar, M. B. Sedra, Int. J. Geom. Methods Mod. Phys. \textbf{12}, 1550017 (2015)

\bibitem{Guna2012}
S. Gunasekaran, R. B. Mann, D. Kubiznak, J. High Energy Phys. \textbf{11}, 110 (2012)

\bibitem{Dutta2013}
S. Dutta, A. Jain, R. Soni, J. High Energy Phys. \textbf{12}, 60 (2013)

\bibitem{Li2014}
G. Q. Li, Phys. Lett. B \textbf{735}, 256 (2014)

\bibitem{Alta2014}
N. Altamirano, D. Kubiznak, R. B. Mann, Z. Sherkatghanad, Galaxies \textbf{2}, 89 (2014)

\bibitem{Liang2016}
J. Liang, C. B. Sun, H. T. Feng, Europhys. Lett. \textbf{113}, 30008 (2016)

\bibitem{Hendi2013}
S. H. Hendi, M. H. Vahidinia, Phys. Rev. D \textbf{88},084045 (2013)

\bibitem{Hendi2016}
S. H. Hendi, S. Panahiyan, B. E. Panah, J. High Energy Phys. \textbf{01}, 129 (2016)

\bibitem{Spa2013}
E. Spallucci, A. Smailagic, Phys. Lett. B \textbf{723}, 436 (2013)

\bibitem{Moa2013}
J. X. Mo, W. B. Liu, Phys. Lett. B \textbf{727}, 336 (2013)

\bibitem{Alta2014b}
N. Altamirano, D. Kubiznak, R. B. Mann, Z. Sherkatghanad, Class. Quant. Grav. \textbf{31}, 042001 (2014)

\bibitem{Sadeghi2014}
J. Sadeghi, H. Farahani, Int. J. Theor. Phys. \textbf{53}, 3683 (2014)

\bibitem{Zhao2013}
R. Zhao, M. Ma, H. Li, L. Zhang 2013 , Adv. High Energy Phys. \textbf{2013}, 371084 (2013)

\bibitem{John2014}
C. V. Johnson, Classical Quant. Grav. \textbf{31}, 205002 (2014)

\bibitem{John2015a}
C. V. Johnson, Classical Quant. Grav. \textbf{33}, 135001 (2016)

\bibitem{John2015b}
C. V. Johnson, arXiv:1511.08782 (2015)

\bibitem{Bel2015b}
A. Belhaj, M. Chabab, H. E. Moumni, K. Masmar, M. B. Sedra, A. Segui, J. High Energy Phys. \textbf{05}, 149 (2015)

\bibitem{Caceres2015}
E. Caceres, P. H. Nguyen, J. F. Pedraza, J. High Energy Phys. \textbf{1509}, 184 (2015)

\bibitem{Sadeghi2015}
J. Sadeghi, K. Jafarzade, arXiv:1504.07744 (2015)

\bibitem{Setare2015}
M. R. Setare, H. Adami, Gen. Relat. Gravity \textbf{47},132 (2015)

\bibitem{John2016}
C. V. Johnson, Entropy \textbf{18}, 120 (2016)

\bibitem{Dolan2011b}
B. P. Dolan, Phys. Rev. D \textbf{84}, 127503 (2011)

\bibitem{Wint}
D. E. Winterbone, \textit{Advanced Thermodynamics for Engineers}, 1st edn. (Butterworth-Heinemann, Oxford, 1997)

\bibitem{Johns2014}
D. C. Johnston, arXiv:1402.1205 (2014)

\bibitem{Vent2001}
S. L. Vent, IJMEE \textbf{29}, 257 (2001)

\bibitem{Golden}
N. Goldenfeld, \textit{Lectures on Phase Transition and the Renormalization Group}, 1st edn. (Westview Press, New York, 1992)


\end{references}

\end{document}